\documentclass[pra,aps,showpacs,showkeys,reprint]{revtex4}
\usepackage{graphicx}
\usepackage{amsmath}
\usepackage{amssymb}

\begin{document}
\title{\bf The minimal length uncertainty and the nonextensive thermodynamics}
\author{Homa Shababi and Pouria Pedram}
\email[Electronic address: ]{p.pedram@srbiau.ac.ir}
\affiliation{Department of Physics, Science and Research Branch,
Islamic Azad University, Tehran, Iran}
\date{\today}

\begin{abstract}
In this paper, we study the thermodynamics of quantum harmonic
oscillator in the Tsallis framework and in the presence of a minimal
length uncertainty. The existence of the minimal length is motivated
by various theories such as string theory, loop quantum gravity, and
black-hole physics. We analytically obtain the partition function,
probability function, internal energy, and the specific heat
capacity of the vibrational quantum system for $1<q<\frac{3}{2}$ and
compare the results with those of Tsallis and Boltzmann-Gibbs
statistics without the minimal length scale.
\end{abstract}

\keywords{Minimal length, Generalized uncertainty principle,
Nonextensive thermodynamics}

\pacs{04.60.Bc}

\maketitle

\section{Introduction}
Various approaches to quantum gravity such as string theory
\cite{Veneziano,Amati,Amati2,Amati3,Amati4,Gross,Konishi},
noncommutative geometry \cite{Capozziello}, and loop quantum gravity
\cite{Garay} motivate the existence of a minimal measurable length.
These theories argue that near the Planck scale, the Heisenberg
Uncertainty Principle (HUP) should be replaced by the so-called
Generalized Uncertainty Principle (GUP) (see
Refs.~\cite{Veneziano,Amati,Amati2,Amati3,Amati4,Gross,Konishi} and
references therein). The effects of gravity play an important role
in the limit of small distances around the Planck length,
$\ell_{Pl}=\sqrt{G \hbar/c^3}\approx 10^{-35}m$, where $G$ is
Newton's constant, $\hbar$ is the Planck constant, and $c$ is the
speed of light.  Equivalently, these effects are dominant in the
limit of high energies near the Planck energy $E_{Pl}=\sqrt{\hbar
c^5/G}\approx 1.956 \times 10^{9} J$.

In nonrelativistic quantum mechanics, a particle with an arbitrary
energy can be accurately localized and its exact location can be
obtained. According to the quantum filed theory and based on
Heisenberg Uncertainty Principle, only particles with infinite
energy can be localized. However, in quantum gravity theories that
try to combine the effects of gravity with quantum mechanics, the
localization of particles completely disappear. In general
relativity, it is possible to access the path of  particles that is
impossible to obtain in quantum mechanics. In fact, in quantum
mechanics determination of exact position of particles requires
infinite momentum uncertainty. Now, in order to achieve the ultimate
theory of quantum gravity, one of the results of combination of
gravity and quantum mechanics is introducing the minimal measurable
length of the order of the Planck length. In this case, particles
lose their point-like description. Thus, to find the position of
particles, a minimal length uncertainty proportional to the Planck
length should be taken into account and the sharp localization
condition for particles is relaxed \cite{Kempf95}.

Moreover, some Gedanken experiments in black-hole physics and
noncommutativity of the spacetime manifold predict the existence of
a minimal observable distance of the order of the Planck length
\cite{Doplicher,Snyder,Connes,Varilly,Douglas}. The generalization
of ordinary uncertainty relation by incorporating quantum gravity
effects requires a modification of the Hamiltonian of various
physical systems which has been studied in the literature
\cite{Fityo,Nozari1,Nozari2,Pedram,Amelino,Rama,Vakili,Abbasiyan}.

According to recent statistical approaches, standard thermodynamics
seems to be not suitable for studying all systems including the
behavior of complex systems which are governed by the nonextensive
statistics \cite{Beck,Zhang,Gupta,Razdan,Tonga}. A nonextensive
statistical mechanics which is proposed by Tsallis \cite{Tsallis88}
and inspired by the probabilistic description of multifractal
geometries, generalizes all thermodynamical quantities such as the
partition function and the internal energy with the entropic index
$q$ that plays an important role  in this framework. Moreover, in
the limit $q\rightarrow1$ it gives the well-known Boltzmann-Gibbs
statistics.

Notice that the nonextensive statistics could be more useful than
the traditional Boltzmann-Gibbs statistics in many situations. The
theory behind the Boltzmann-Gibbs entropy implies a strong
dependence on initial conditions. However, most materials behave
quite independently of the initial conditions. In other words, the
usage of the well-known BG statistical mechanics occasionally fails
and leads to serious difficulties. In this regards, we can mention
long-range interactions, e.g., $d = 3$ gravitation \cite{Saslaw},
long-range microscopic memory, e.g., nonmarkovian stochastic
processes \cite{Risken,Caceres}, and conservative or dissipative
systems multifractal-like structures \cite{Tsallis99}. More examples
include granular systems \cite{Taguchi}, Levy anomalous diffusion
\cite{Montroll}, pure-electron plasma two-dimensional turbulence
\cite{Huang}, phonon-electron anomalous thermalization in
ion-bombarded solids \cite{Koponen}, solar neutrinos \cite{Clayton},
peculiar velocities of galaxies \cite{Bahcall}, and inverse
bremsstrahlung in plasma \cite{Liu94} and black-holes \cite{Maddox}.

During the last decade, Tsallis statistics has attracted much
attention in the literature
\cite{Feng,Liyan,AragaoRego,Carrete,Fukuda,Hasagawa,Chakrabarti,Tsallis98,Lenzi,E.KLenzi}.
Among these investigations, we can mention the harmonic oscillator
in both classical and quantum domains \cite{Tsallis98}, the phase
space of Tsallis entropy for the harmonic oscillator \cite{Sadeghi},
the quantum entanglement of maximum Tsallis entropy state
\cite{Abe}, path integral approach to the nonextensive canonical
density matrix \cite{E.K.Lenzi2000}, the hydrogen atom in three
dimensions \cite{Ghosh}, and the sensitivity of the population state
to the value of $q$ for a two-state system \cite{Nasimi}. Moreover,
Tsallis entropy is compared to two different entropies, namely,
Renyi entropy $S_{q}^R=\frac{1}{1-q}\ln\sum_{n}(p_{n})^q$, and the
normalized Tsallis entropy,
$S_{q}^{NT}=\frac{1}{1-q}\left(1-\frac{1}{\sum_{n}p_{n}^q}\right)$,
which is shown that the Tsallis entropy is more stable
\cite{SumiyoshiAbe}. Recently, some authors have investigated the
exact vibrational partition function for quantum nonextensive
harmonic oscillator \cite{Keshavarzi}. They showed that for all $T$
and $1<q<2$ the inequality $Z>Z^{*}$ holds and $Z-Z^{*}$ increases
as $q$ and $T$ grow up where $Z$ and $T$ represent the partition
function and temperature, respectively. Here and throughout the
paper the symbol $*$ denotes the quantities obtained in the
extensive statistics, i.e., Boltzmann-Gibbs thermodynamics.

Nowadays, it is believed that the minimal length scale naturally
appears in various areas of high energy physics \cite{Garay}. In
particular, the effects of minimal length scale on thermodynamics of
various physical systems have been studied in the literature (see,
e.g.,
Refs.~\cite{Fityo,Nozari1,Nozari2,Pedram,Amelino,Rama,Vakili,Abbasiyan}).
At this point, it is worth explaining why the idea of minimal length
scale can be incorporated in thermodynamics. Consider a system which
cannot be probed to distances smaller than some length scale. This
phenomenon can be an intrinsic property of the system such as the
minimal length scale in high energy physics which is of the order of
the Planck length
\cite{Amati,Gross,Garay,Kempf95,Mead,Maggiore,Witten,Scardigli,Szabo}.
On the other hand, it can be due to the size effect of the
measurement apparatus. For instance, in finance as a complex system,
the existence of a minimal length scale is necessary due to the fact
that the stock price is indeed a discrete variable \cite{P.Pedram3}.
In thermodynamics, the gas particles have a nonzero size of the
order of the atom's radius. So, these particles cannot probe
distances smaller than their finite size which is indeed the minimal
length scale of the system. Here, we use a framework that is
compatible with this minimal length scale.

In this paper, we study the thermodynamics of the quantum harmonic
oscillator in the presence of a minimal length and in the
nonextensive Tsallis statistics. Although, other forms of GUP which
imply both minimal measurable length and maximal momentum could be
considered \cite{Magneijo,Smolin,Cortes,A.F,E.C}, to obtain exact
solutions we only consider the existence of a minimal length. Notice
that at ordinary distances the effects of gravity are not dominant
and the ordinary physical laws can be used without difficulty.
However, at small distances of the order of Planck length, the
effects of gravity are considerable and the modification of physical
laws would be required. The usage of the Tsallis statistics is based
on the fact that physical systems with long-range interactions,
long-time memories, and multifractal structures can hardly be
treated within the traditional BG statistical framework. Therefore,
the application of nonextensive statistical models is unavoidable
for these systems.

The aim of our investigation is to analyze the behavior of a
nonextensive thermodynamical system in the framework of the quantum
gravity. Also, the modification of the thermodynamical variables in
the presence of the minimal length uncertainty is addressed. The
nonextensive statistical model contains the effect of multifractal
geometry and therefore the concept of discrete geometric
constructions can be essentially derived from multifractal geometry.
However, the probabilities in nonextensive models depend on the
energy spectrum of the system. In fact, the statistical model does
not obtain the energy spectrum, it only use them to construct the
partition function. So, when an statistical model agrees with the
concept of discrete geometric constructions, the energy spectrum
should be also obtained from a framework which agrees with this
concept. In other words, a consistent way to find the nonextensive
partition function is to use the GUP-corrected energy spectrum as
the input of the nonextensive model which both imply the discrete
geometry.

Here, the vibrational partition function, probability function,
internal energy, and the specific heat capacity  are obtained for
$1<q<\frac{3}{2}$. We also compare our results with the
corresponding quantities in the absence of minimal length
\cite{Keshavarzi} and with the Boltzmann-Gibbs thermodynamics. Note
that the black-body radiation in Tsallis statistics and in the
presence of the minimal length can also be treated using this model.

\section{Quantum oscillator in the presence of a minimal length scale}
In ordinary quantum mechanics, Heisenberg uncertainty principle
allows us to measure the position and momentum of a particle
separately with an arbitrary precision. But in the presence of a
minimal value for the position measurement, the Heisenberg
uncertainty relation should be modified. One way to incorporate the
minimal length is to generalize the Heisenberg uncertainty principle
as follows \cite{Kempf95}
\begin{eqnarray}
\Delta X \Delta P\geq \frac{\hbar}{2}\left(1+\beta
\left[\left(\Delta P\right)^{2}+ {\left\langle P\right\rangle
}^2\right]\right),
\end{eqnarray}
where $\beta$ is the deformation parameter and defined as
$\beta=\beta_{0}/(M_{Pl}c)^2$ which $M_{Pl}$ is the Planck mass and
$\beta_{0}$ is of the order of unity. To show this, note that the
absolute minimal measurable length is obtained by saturating the
above inequality. So, we find $(\Delta
X)_{min}=\hbar\sqrt\beta\sqrt{1+\beta{\left\langle P\right\rangle
}^2}$ and the absolute minimal measurable length is given by
$(\Delta X)_{min}=\hbar\sqrt\beta$ for $\left\langle
P\right\rangle=0$. In the context of string theory this length can
be interpreted as the length of the string and therefore the
string's length is proportional to the square root of the GUP
parameter. In thermodynamics, this minimal length is given by the
size of the atoms where distances smaller than $(\Delta X)_{min}$
cannot be probed by the particles. Notice that, since we have
$(\Delta X)_{min}=\sqrt{\beta_0}\ell_{Pl}$, the minimal length scale
is not essentially equal to the Planck length, but it is of the
order of the Planck length. So, $(\Delta X)_{min}$ could be smaller
or bigger than the Planck length upon choosing $\beta_0<1$ or
$\beta_0>1$. The exact value for $\beta_0$ is hoped to be obtained
in the future experiments. In one dimension, the generalized
uncertainty relation leads to the following deformed commutation
relation
\begin{eqnarray}
\left [X,P\right]=i\hbar\left(1+\beta P^{2}\right).
\end{eqnarray}
According to Refs.~\cite{Pedram,Pedram1}, $X$ and $P$ can be written
in terms of ordinary position and momentum operators as
\begin{eqnarray}\label{99}
X=x,\hspace{2cm}P=\frac{\tan(\sqrt{\beta}p)}{\sqrt{\beta}},
\end{eqnarray}
where $[x,p]=i\hbar$. Moreover, all Hamiltonians will be modified
due to the deformed commutation relation. In this \emph{formally}
self-adjoint representation, the Hamiltonian
$H=\frac{P^{2}}{2m}+V(X)$ can be expressed as
\begin{eqnarray}
H=\frac{\tan^{2}{(\sqrt{\beta}p)}}{2\beta m}+V(x).
\end{eqnarray}

Here, we study the thermodynamics of the quantum harmonic oscillator
which its Hamiltonian is given by
$H=\frac{P^2}{2m}+\frac{1}{2}m\omega^2X^2$. So, using the
representation (\ref{99}), the generalized Schr\"odinger equation in
momentum space reads
\begin{eqnarray}
-\frac{1}{2}m\hbar^2\omega^2\frac{d^2\phi(p)}{dp^2}+\frac{\tan^2\left(\sqrt{\beta}p\right)}{2m\beta}\phi(p)=E\phi(p).
\end{eqnarray}
Using the new variable $\xi=\sqrt{\beta}p$, the above equation can
be written as
\begin{eqnarray}
\frac{d^2\phi(\xi)}{d\xi^2}+\left(\epsilon-\frac{V}{\cos^2\xi}\right)\phi(\xi)=0,
\end{eqnarray}
where $V=\left(m\beta\hbar\omega\right)^{-2}$ and
$\epsilon=V(1+2m\beta E)$. Now, taking
\begin{eqnarray}
\phi_n(\xi)=P_n(s)\cos^\lambda \xi,
\end{eqnarray}
leads to
\begin{eqnarray}
(1-s^2)\frac{d^2P_n(s)}{ds^2}-s(1+2\lambda)\frac{dP_n(s)}{ds}+(\epsilon-\lambda^2)P_n(s)=0,
\end{eqnarray}
where $s=\sin\xi$ and
\begin{eqnarray}
V=\lambda(\lambda-1),\hspace{2cm}\lambda=\frac{1}{2}\left[1+\sqrt{1+\frac{4}{m^2\beta^2\hbar^2\omega^2}}\,\right].
\end{eqnarray}
The solutions of the above equation for $\epsilon=(n+\lambda)^2$ are
given by the Gegenbauer polynomials $C^{\lambda}_n(s)$. Therefore,
the exact solutions read
\begin{eqnarray}\label{solg}
\phi_n(p)&=&N_n\,C^{\lambda}_n\left(\sin(\sqrt{\beta}p)\right)\cos^\lambda
(\sqrt{\beta}p),\\
E_n&=&\hbar\omega\left(n+\frac{1}{2}\right)\left(\sqrt{1+\gamma^2/4}+\gamma/2\right)+\frac{1}{2}\hbar\omega\gamma
n^2,\label{0}
\end{eqnarray}
for $n=0,1,2,\ldots$. Here $\gamma=m\beta\hbar\omega$, $N_n$ is the
normalization coefficient, and the Gegenbauer polynomials are
defined as \cite{Gradshteyn}
\begin{eqnarray}
C^{\lambda}_n(s)=\sum_{k=0}^{[n/2]}(-1)^k\frac{\Gamma(n-k+\lambda)}{\Gamma(\lambda)k!(n-2k)!}(2s)^{n-2k}.
\end{eqnarray}
Note that in the absence of minimal length scale
($\beta\sim\gamma\rightarrow0$), the ordinary energy spectrum of the
quantum harmonic oscillator is obtained, i.e.,
$E_n=\hbar\omega\left(n+\frac{1}{2}\right)$.

\section{Thermodynamics in the Tsallis framework}
The entropy of Tsallis statistics is given by \cite{Tsallis88}
\begin{eqnarray}\label{1}
S_{q}=k\frac{1-\sum_{n} p_{n}^{q}}{q-1},
\end{eqnarray}
where $p_{n}$ is the probability of the \emph{n}th microstate, $k$
is a positive constant, and $q$ is the entropic index. Using two
constrains $\sum_{n=1}^{w}p_{n}=1$  and
$\sum_{n=1}^{w}E_{n}p_{n}^{q}=U_{q}$ that maximize Eq.~(\ref{1})
where $w$ is the total number of microstates of the system
\cite{Tsallis88,Tsallis91}, the generalized statistical probability
$p_{n}$ and the partition function $Z_{q}$  are given by
\begin{eqnarray}\label{2}
p_{n}&=&\frac{\left[1-b(1-q)E_{n}\right]^{\frac{1}{1-q}}}{Z_{q}},\\
Z_{q}&=&\sum_{n=0}^{\infty}\left[1-b(1-q)E_{n}\right]^{\frac{1}{1-q}},\label{3}
\end{eqnarray}
where $b^{-1}=kT$ is the Lagrange parameter related to the internal
energy $ U_{q}$, and $E_{n}$ is the energy of the \emph{n}th quantum
state. Now, using  Eq.~(\ref{0}) up to the first order of $\gamma$,
the vibrational partition function reads
\begin{eqnarray}\label{5}
Z_{vib,q}=\sum_{n=0}^{\infty}\frac{1}{\left[1+b(q-1)(n+\frac{1}{2})\hbar\omega(1+\frac{\gamma}{2})+\frac{b\hbar\omega\gamma(q-1)}{2}n^2\right]^{\frac{1}{q-1}}},
\end{eqnarray}
where $n$ is the quantum number and $\omega$ is the frequency of the
harmonic oscillator.

Notice that, the convergence of the above equation implies $q>1$.
Now, since $\gamma$ is a small parameter, up to the first order we
have
\begin{eqnarray}\label{7}
Z_{vib,q}=[b(q-1)\hbar\omega]^{\frac{1}{1-q}}\sum_{n=0}^{\infty}\frac{1}{(n+a)^s}\left[1-\frac{\gamma
s}{2}\left(\frac{n^2+n+\frac{1}{2}}{n+a}\right)\right],
\end{eqnarray}
where $s=\frac{1}{q-1}$ and
$a=\frac{1}{b(q-1)\hbar\omega}+\frac{1}{2}$. Consequently, using the
Hurwits zeta function
$\zeta(s,a)=\sum_{n=0}^{\infty}\frac{1}{(n+a)^s}$ and after some
algebra, the analytic relation for the partition function is given
by
\begin{eqnarray}\label{9}
Z_{vib,q}&=&Z_{0}+\gamma
s[b(q-1)\hbar\omega]^{\frac{1}{1-q}}\left\{\left(a-\frac{1}{2}\right)\zeta(s,a)-\frac{1}{2}\zeta(s-1,a)-
\left(\frac{a^2}{2}-\frac{a}{2}+\frac{1}{4}\right)\zeta(s+1,a)\right\},
\end{eqnarray}
which $Z_{0}\equiv Z_{vib,q}(\gamma=0=\beta)$ is the vibrational
partition function of quantum harmonic oscillator in the absence of
minimal length scale \cite{Keshavarzi}, namely
$Z_{0}=[b(q-1)\hbar\omega]^{\frac{1}{1-q}}\zeta(s,a)$. Taking into
account two conditions imposed by the definition of Hurwits zeta
function, i.e., $s>1$ and $a>0$, the validity of Eq.~(\ref{9})
implies the inequality $1<q<\frac{3}{2}$. Also,  in the limit of
$q\rightarrow1$, Eq.~(\ref{9}) gives the corresponding
Boltzmann-Gibbs partition function in the presence of the minimal
length scale \cite{Pedram}. For our case, the nonextensive
probability function can be expressed as
\begin{eqnarray}\label{11}
p_{n}=\frac{\left[1+b(q-1)(n+\frac{1}{2})\hbar\omega(1+\frac{\gamma}{2})+\frac{b\hbar\omega\gamma(q-1)}{2}n^2\right]^{\frac{1}{1-q}}}{Z_{vib,q}}.
\end{eqnarray}

\begin{figure}
\centering
\includegraphics[width=11cm]{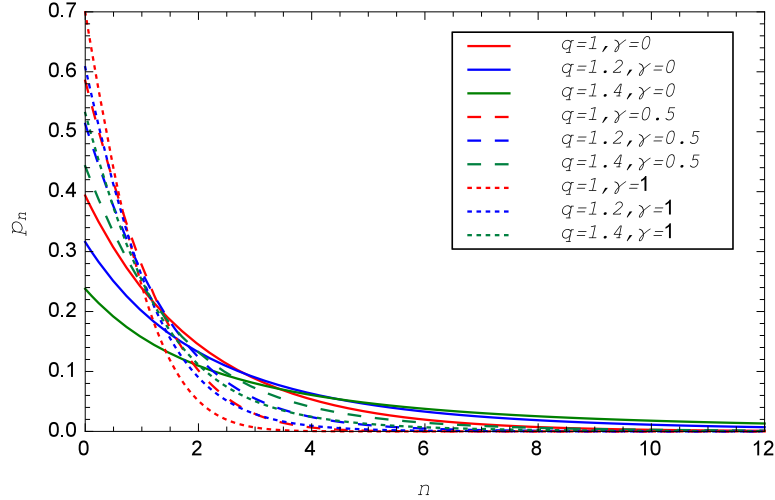}
\caption{\label{fig1}Comparison between $p_{n},p_{n}^{*}$ and
$p_{n}(\gamma=0),p_{n}^{*}(\gamma=0)$ versus their energy states for
$1<q<1.5$ and $kT/\hbar\omega=2$.}
\end{figure}

\begin{figure}
\centering
\includegraphics[width=11cm]{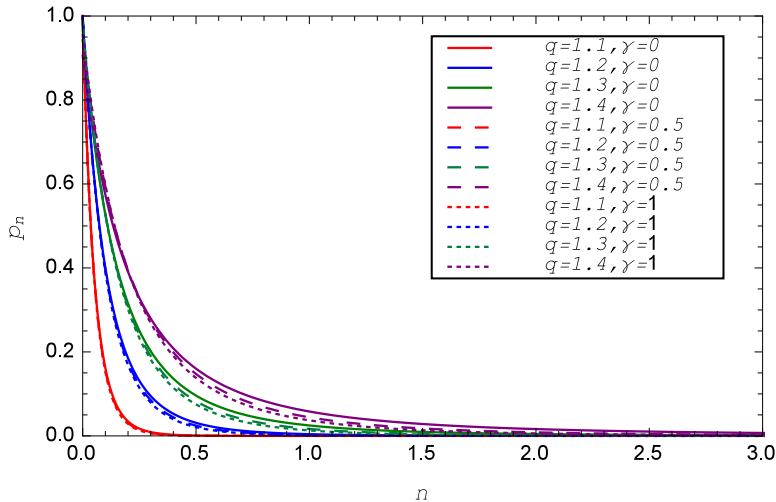}
\caption{\label{fig2}Comparison between $p_{n}$ and
$p_{n}(\gamma=0)$ versus their energy states for $1<q<1.5$ and
$kT/\hbar\omega\rightarrow0$.}
\end{figure}

\begin{figure}
\centering
\includegraphics[width=11cm]{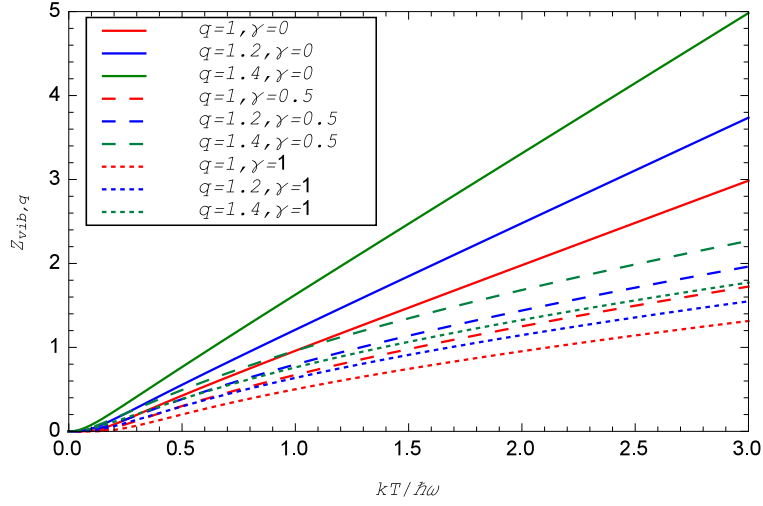}
\caption{\label{fig3}Comparison between  partition functions of
extensive and nonextensive quantum harmonic oscillators in the
presence of a minimal length scale, ${Z}_{vib,q}$, and their
corresponding quantum harmonic oscillators without the effects of
minimal length scale, ${Z}_{vib,q}(\gamma=0)$, versus their
temperatures.}
\end{figure}

To investigate the results, we plotted the probability functions
versus the quantum number for two temperatures in Figs.~\ref{fig1}
and \ref{fig2}. For $T=2$, as it is shown in Fig.~\ref{fig1}, we
have $p_{n}^{*}>p_{n}$ for small $n$ and $p_{n}>p_{n}^{*}$ for large
$n$. Moreover, these results can be compared with those in the
absence of the minimal length scale \cite{Keshavarzi}. Indeed, for
small quantum numbers we have $p_{n}>p_{n}(\gamma=0)$ and
$p_{n}^{*}>p_{n}^{*}(\gamma=0)$. However, this result becomes
reversed for large $n$. It is shown that at a given temperature and
$q$, when $\gamma$ increases, the probability functions are also
increase for small $n$. But for large $n$ this behavior  becomes
reverse. As it is known, in zero temperature limit and unlike the BG
statistics, the probability function is not equal to unity for the
ground state and does not vanish for higher energy states. This
behavior is depicted in Fig.~\ref{fig2}. Thus, for $T\rightarrow0$
we find $p_{n}(\gamma=0)>p_{n}$ and $p_{n}^{*}(\gamma=0)>p_{n}^{*}$
for all $n$. Based on Eq.~(\ref{9}), the partition function depends
on the temperature, the entropic index $q$, and $\gamma$. As it is
shown in Fig.~\ref{fig3}, we obtain ${Z}_{vib,{q}'}>{Z}_{vib,q}$ for
${{q}'}>q$ and all $T$. Also, we have
${Z}_{vib,q}<Z_{vib,q}(\gamma=0)$ and
${{Z}^{*}}_{vib,q}<Z^{*}_{vib,q}(\gamma=0)$. For
$\gamma_{1}>\gamma_{2}>\gamma_{3}$, it is concluded that for fixed
entropic indices and temperatures, we have
${Z}_{vib,q}(\gamma_{1})<{Z}_{vib,q}(\gamma_{2})<{Z}_{vib,q}(\gamma_{3})$.

Using the properties of the Hurwits zeta function and the definition
of the internal energy in the  Tsallis statistics
\begin{eqnarray}\label{12}
U=-\frac{\partial}{\partial b}
\left[\frac{(Z_{q})^{1-q}-1}{1-q}\right]=-\frac{1}{(Z_{q})^{q}}
\left(\frac{\partial Z_{q}}{\partial b}\right),
\end{eqnarray}
we find
\begin{eqnarray}
U&=&U_{0}+\frac{\gamma s \hbar \omega
[b(q-1)\hbar\omega]^{\frac{q}{1-q}}}{Z_{q}^{q}}\left\{\left(a-\frac{1}{2}\right)\zeta(s,a)-\frac{1}{2}\zeta(s-1,a)-\left(\frac{a^{2}}{2}-\frac{a}{2}+\frac{1}{4}\right)\zeta(s+1,a)\right\}\nonumber\\
 && -\frac{\gamma
s[b(q-1)\hbar\omega]^{\frac{1}{1-q}}}{Z_{q}^{q}}
\times
\Bigg\{\left[\frac{1}{b^3(q-1)^2\hbar^2\omega^2}+\frac{(a-\frac{1}{2})s}{b^{2}(q-1)\hbar\omega}\right]\zeta(s+1,a)-\frac{(s+1)}{2b^{2}(q-1)\hbar\omega}\zeta(s,a)\nonumber\\
 && +\frac{(-\frac{a^2}{2}+\frac{a}{2}-
\frac{1}{4})(s+1)}{b^{2}(q-1)\hbar\omega}\zeta(s+2,a)\Bigg\},\label{13}
\end{eqnarray}
where $U_{0}\equiv U(\gamma=0) $ is the internal energy corresponds
to the absence of the minimal length scale which is given by
\cite{Keshavarzi}
\begin{eqnarray}\label{14}
U_{0}=\frac{\hbar \omega[(q-1)b\hbar
\omega]^{\frac{q}{1-q}}\zeta(s,a)-[(q-1)b\hbar
\omega]^{\frac{1}{1-q}}\frac{1}{(q-1)^2 b^2 \hbar
\omega}\zeta(s+1,a)}{Z_{0}^{q}}.
\end{eqnarray}

In contrast to BG statistics, because of the kind of the
distribution and the definition for internal energy, in the
nonextensive model the internal energy at zero temperature will not
be equal to the ground state energy \cite{Keshavarzi}. Also, when
the temperature goes to zero, the heat capacity for extensive and
nonextensive systems approaches zero but with different slopes. In
nonextensive systems we observe an increased energy fluctuation
which is due to the increasing the number of accessible states in
comparison with extensive ones. The difference in the number of
accessible states will be dominated as temperature goes to zero
\cite{Keshavarzi}.

\begin{figure}
\centering
\includegraphics[width=11cm]{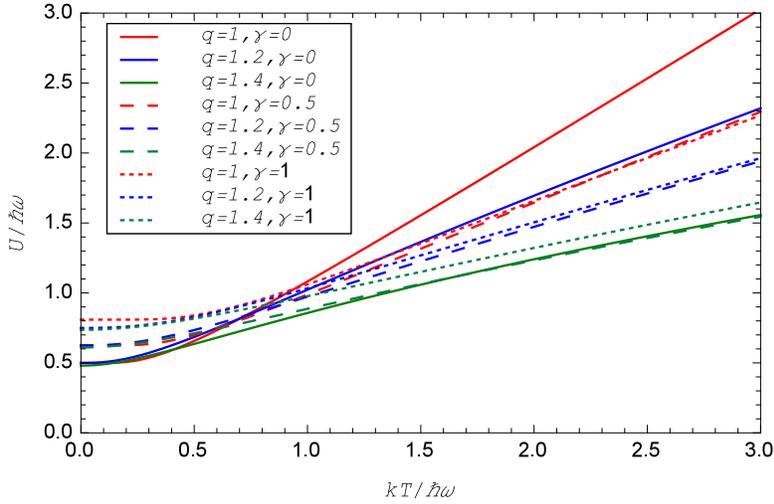}
\caption{\label{fig4}Comparison between internal energy of extensive
and nonextensive quantum harmonic oscillator in the presence of a
minimal length scale, $U^{*},U$ and its corresponding quantum
harmonic oscillator without the effects of minimal length scale,
$U^{*}(\gamma=0),U(\gamma=0)$, versus their temperatures.}
\end{figure}

As it is shown in Fig.~\ref{fig4}, for low temperatures we have
$U>U^{*}$, but for high temperatures the situation is revered,
namely $U^{*}>U$. Note that, this result is not unexpected. In fact,
as it is mentioned before, at zero temperature and in
Boltzmann-Gibbs statistics for both $\gamma=0$ and $\gamma\ne0$, the
probability of ground state is equal to $1$ and it is zero for other
excited states. For the nonextensive case, the fluctuations of
energy increases and this is because of the increasing of the
occupation number of states. In the presence of minimal length, in
addition to the fluctuations of the nonextensive statistics, we have
fluctuations due to the minimal length scale. Indeed, in
nonextensive statistics for both $\gamma = 0$ and $ \gamma\neq 0$,
the probability of ground state is less than $1$ for
$T\rightarrow0$. So, in the limit of low temperatures, the
nonextensive internal energy must be larger than its corresponding
quantity in the absence of this length scale and for the high
temperatures the situation becomes reversed. Also, the behavior of
internal energy is manifest in this figure for three different
values of $\gamma$. For low temperatures, if
$\gamma_{1}>\gamma_{2}>\gamma_{3}$ it is concluded that for fixed
entropic indices and temperatures,
$U(\gamma_{1})>U(\gamma_{2})>U(\gamma_{1})$, but for high
temperatures, we obtain different results upon choosing $q$. The
reason for the reduction of internal energy in comparing to $\gamma
= 0$ is a consequence of the reduction of phase space volume due to
possible definition of a rescaled $\hbar$. In fact, in the presence
of the minimal length uncertainty, the volume of the fundamental
cell increases and the number of degrees of freedom reduces
consequently \cite{Nozari2}. Notice that, in the limit of
$q\rightarrow1$, we obtain the extensive internal energy for quantum
harmonic oscillator in the presence of minimal length scale
\cite{Pedram}.

\begin{figure}[h]
\centering
\includegraphics[width=11cm]{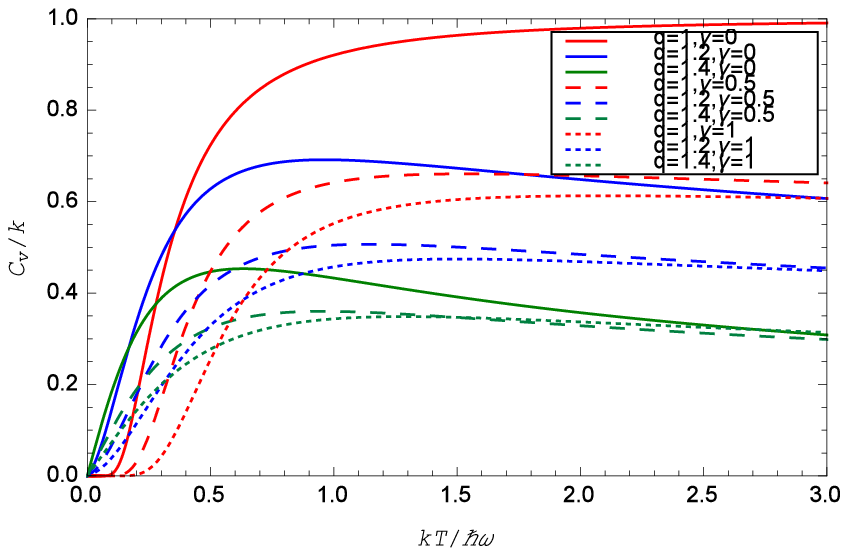}
\caption{\label{fig5}Comparison between heat capacity of extensive
and nonextensive quantum harmonic oscillator in the presence of a
minimal length scale, $C_{V}^{*},C_{V}$ and its corresponding
quantum harmonic oscillator without the effects of minimal length
scale, $C^{*}_{V}(\gamma=0),C_{V}(\gamma=0)$, versus their
temperatures.}
\end{figure}

Another useful thermodynamical quantity is the specific heat
capacity. As it is shown in Fig.~\ref{fig5}, in the limit of low
temperatures we observe $C_{V}>C_{V}^{*}$. This is due to the fact
that at low temperatures for the nonextensive case, the number of
accessible states are modified which results in the increase of the
heat capacity. However, in the limit of high temperatures we have
$C_{V}<C_{V}^{*}$. By comparing the heat capacity in the presence of
minimal length $C_{V}$, and its corresponding value in the absence
of this minimal length $C_{V}(\gamma=0)$, we find
$C_{V}^{*}<C_{V}^{*}(\gamma=0)$ and $C_{V}<C_{V}(\gamma=0)$. Also,
for $\gamma_{1}>\gamma_{2}>\gamma_{3}$ we have
$C_{V}(\gamma_{1})<C_{V}(\gamma_{2})<C_{V}(\gamma_{3})$. In the
nonextensive case unlike the BG case, the heat capacity represents a
peak which is also observed for two-level systems \cite{Tsallis98}.
As it is shown in Fig.~\ref{fig5}, $C_{V}^{*}-C_V$ increases as $q$
grows up. This behavior is due to the relationship between the heat
capacity and the internal energy fluctuations \cite{Liyan}. In the
limits of $q\rightarrow1$ and $\gamma\rightarrow0$ the BG heat
capacity for the ordinary quantum harmonic oscillator is recovered.

\section{Conclusions}
In this paper, we have studied the thermodynamics of quantum
harmonic oscillator in the presence of a minimal length uncertainty
and in the Tsallis framework. The notion of a nonzero minimal length
scale is motivated by various phenomena in high energy physics such
as string theory, loop quantum gravity, and black-hole physics. In
fact, in the limit of high energies near the Planck energy or
equivalently in the limit of small distances near the Planck length,
the effects of gravity become dominant which modifies the physical
laws. On the other hand the importance of the Tsallis statistics as
a nonextensive statistical model is due to the fact that the
extensive BG statistics has a strong dependence on the initial
conditions which is not hold for the most materials. So by
considering this two effects, we study the vibrational partition
function, probability function, internal energy, and the heat
capacity analytically for $ 1<q<\frac{3}{2}$. It is shown that the
nonextensive internal energy is larger than its extensive
counterpart for low temperatures and vise versa for the high
temperatures. This behavior is related to the different definitions
of the internal energy in the extensive and nonextensive statistics
and the dependence of the probability function to the entropic
index. It is also shown that $\frac{\partial{ Z_{vib,q}}}{\partial
T}>0$ and ${Z}_{vib,{q}'}>{Z}_{vib,q}$ for ${{q}'}>q$. Then, the
heat capacity is studied and it is indicated that at high
temperatures the nonextensive heat capacity is smaller than the
extensive one and this is because of the statistical correlations in
the nonextensive statistics. At low temperatures, due to the number
of accessible states a larger fluctuation is expected. We also
studied the behavior of the probability function in the presence of
a minimal length scale  for two different temperatures and found
that for the nonextensive statistics, the probability of finding the
system at the ground state is not equal to unity even at the zero
temperature. Finally, in the limit of $q\rightarrow1$, the
corresponding Boltzmann-Gibbs results in the presence of the minimal
length uncertainty is recovered. These results could also shed
lights on the black-body radiation with minimal length and in
Tsallis thermodynamics.

\end{document}